\documentclass[aps,prl,twocolumn,groupedaddress]{revtex4}

\usepackage{graphicx}
\usepackage{dcolumn}
\usepackage{bm}
\usepackage{xcolor}

\begin{document}
\title{Metal-insulator-like transition, superconducting dome and topological electronic structure in Ga-doped Re$_{3}$Ge$_{7}$}
\author{Yanwei Cui$^{1,2,3}$\footnote{These authors contributed equally to this work.}}
\author{Siqi Wu$^{3*}$}
\author{Qinqing Zhu$^{1,2,4}$}
\author{Guorui Xiao$^{1,2,3}$}
\author{Bin Liu$^{1,2,4}$}
\author{Jifeng Wu$^{1,2,4}$}
\author{Guanghan Cao$^{3}$}
\author{Zhi Ren$^{1,2}$\footnote{Email: renzhi@westlake.edu.cn}}

\affiliation{$^{1}$Key Laboratory for Quantum Materials of Zhejiang Province, School of Science, Westlake University, 18 Shilongshan Road, Hangzhou 310024, P. R. China}
\affiliation{$^{2}$Institute of Natural Sciences, Westlake Institute for Advanced Study, 18 Shilongshan Road, Hangzhou 310024, P. R. China}
\affiliation{$^{3}$Department of Physics, Zhejiang University, Hangzhou 310027, P. R. China}
\affiliation{$^{4}$Department of Physics, Fudan University, Shanghai 200433, P. R. China}

\begin{abstract}
Superconductivity frequently appears by doping compounds that show a collective phase transition.
So far, however, this has not been observed in topological materials.
Here we report the discovery of superconductivity induced by Ga doping in orthorhombic Re$_{3}$Ge$_{7}$, which undergoes a second-order metal-insulator-like transition at $\sim$58 K and is predicted to have a nontrivial band topology. It is found that the substitution of Ga for Ge leads to hole doping in Re$_{3}$Ge$_{7-x}$Ga$_{x}$. As a consequence, the phase transition is gradually suppressed and disappears above $x$ = 0.2. At this $x$ value, superconductivity emerges and $T_{\rm c}$ exhibits a dome-like doping dependence with a maximum value of 3.37 K at $x$ = 0.25. First-principles calculations suggest that the phase transition in Re$_{3}$Ge$_{7}$ is associated with an electronic instability driven by Fermi surface nesting and the nontrival band topology is preserved after Ga doping.
Our results indicate that Ga-doped Re$_{3}$Ge$_{7}$ provides a rare opportunity to study the interplay between superconductivity and competing electronic states in a topologically nontrivial system.
\end{abstract}

\maketitle
\maketitle
\section{I. Introduction}
The emergence of superconductivity (SC) in the vicinity of a competing electronic state has received great attention over the past few decades. Well-known examples include cuprates \cite{cuprate}, heavy fermions \cite{HF}, transition metal chalcogenides \cite{CDW}, and iron pnictides \cite{ironpnictide}. Their parent compounds exhibit either an antiferromagnetic \cite{parent1,parent2,parent3} or a charge-density-wave (CDW) transition \cite{parent4}, which is sometimes accompanied by a change from metallic to nonmetallic behavior \cite{MT1,MT2}. Understanding the interplay between these transitions and SC not only sheds light on the pairing mechanism, but also provides an effective route for the search of new superconductors. Recently, the topological aspects of superconductors have become a focus of interest \cite{TSCreview1,TSCreview2,TSCreview3}. These topological superconductors (TSCs) have a bulk superconducting gap but host gapless edge states consisting of Majorana fermions. Many proposals to create TSCs have been put forward and a significant progress has been made. In particular, chemical doping of topological materials is shown to be one of the most promising ways for this purpose \cite{doping1,doping2,doping3,doping4,doping5,doping6,doping7}. However, in all known cases, there exit no competing ground state in the parent compounds.

Re$_{3}$Ge$_{7}$ is the only binary phase in the Re-Ge system and crystallizes in the orthorhombic structure with the $Cmcm$ space group \cite{Re3Ge7}, which is sketched in Figs. 1(a) and (b).
Its structure can be viewed as consisting of isolated Re$_{3}$B-type ReGe$_{3}$ and double NbAs$_{2}$-type Re$_{2}$Ge$_{4}$ prisms \cite{Re3Ge7}. Although Re$_{3}$Ge$_{7}$ has been known to exist for nearly 40 years, its thermodynamic and transport properties are reported only very recently \cite{Re3Ge7PRM}. The results indicate that the compound is weakly diamagnetic while undergoes a second-order phase transition below 58.5 K as confirmed by specific heat measurements. This transition is accompanied by a drop in the diamagnetic susceptibility, a metal-to-insulator-like transition in resistivity, and a strong reduction in electron carrier concentration. While it is suspected that the phase transition in Re$_{3}$Ge$_{7}$ has a structural origin, no low-temperature x-ray study has been performed. On the other hand, theoretical calculations suggest that Re$_{3}$Ge$_{7}$ is a high symmetry point topological semimetal \cite{Re3Ge7topology1}. Since the combination of these properties in a single material is uncommon, it is of significant interest to see what would be the ground state once the phase transition is suppressed.

\begin{figure*}
\includegraphics*[width=17.3cm]{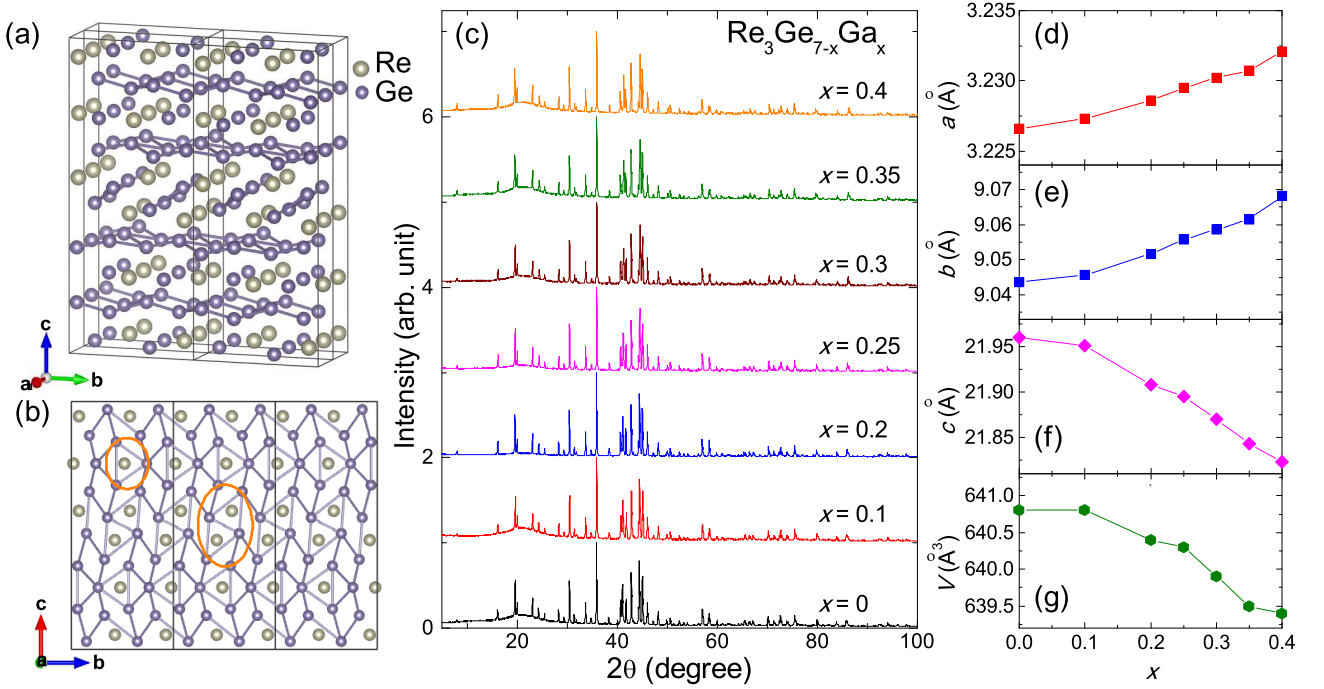}
\caption{
(a) A three-dimensional view of the orthorhombic structure of Re$_{3}$Ge$_{7}$.
(b) Structure of Re$_{3}$Ge$_{7}$ projected perpendicular to the $a$-axis. The circles indicate the ReGe$_{3}$ and Re$_{2}$Ge$_{4}$ building blocks.
(c) Room temperature powder XRD patterns for the series of Re$_{3}$Ge$_{7-x}$Ga$_{x}$ samples.
(d-g) Variation of $a$-, $b$-, $c$-axis lattice parameters and unit-cell volume of Re$_{3}$Ge$_{7-x}$Ga$_{x}$ as a function of the Ga content $x$.
}
\label{fig1}
\end{figure*}

 \begin{figure}
\includegraphics*[width=7.5cm]{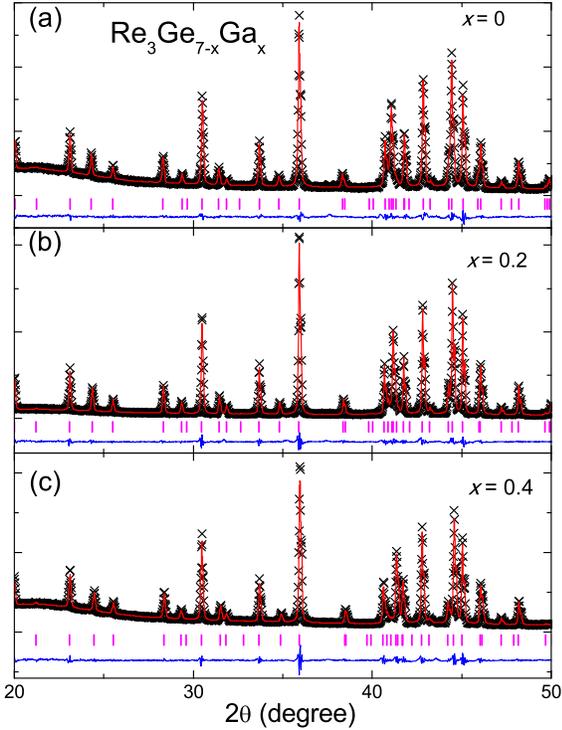}
\caption{
(a-c) Structural refinement profiles in the 2$\theta$ range of 20 to 50$^{\circ}$ for the Re$_{3}$Ge$_{7-x}$Ga$_{x}$ samples with $x$ = 0, 0.2 and 0.4, respectively.
}
\label{fig1}
\end{figure}

Here we show that, upon hole doping by substituting Ge with Ga, Re$_{3}$Ge$_{7}$ turns into a bulk superconductor. The resulting temperature-doping phase diagram for Re$_{3}$Ge$_{7-x}$Ga$_{x}$ resembles closely those of the correlated electron systems. Especially, as the metal-insulator transition is suppressed continuously and disappears completely for $x$ $>$ 0.2, a dome-like superconducting phase is observed for $x$ between 0.2 and 0.4 with a maximum $T_{\rm c}$ of 3.37 K at $x$ = 0.25. The effect of Ga doping on the electronic band structure and topology in Re$_{3}$Ge$_{7}$ is investigated by first-principles calculations, and the implication of these results is discussed.
\section{II. Results and Discussion}
The XRD results for the series of Re$_{3}$Ge$_{7-x}$Ga$_{x}$ samples at room temperature are displayed in Fig. 1(c).
The patterns are very similar and all the diffraction peaks can be well indexed on the basis of an orthorhombic unit-cell with the $Cmcm$ space group.
The refined lattice parameters as well as the unit-cell volume $V$ are plotted as a function of the Ga content $x$ in Figs. 1(d)-(g).
For undoped Re$_{3}$Ge$_{7}$ ($x$ = 0), the $a$-, $b$- and $c$-axis lattice constants are found to be 3.227(1) {\AA}, 9.044(1) {\AA}, and 21.960(1) {\AA}, respectively, which are in excellent agreement with the previous report \cite{Re3Ge7}.
With increasing $x$, both $a$- and $b$-axis expand, which is as expected since since the atomic radius of Ga (1.388 {\AA}) is larger than that of Ge (1.349 {\AA}) \cite{radius}.
Nevertheless, the $c$-axis shrinks more rapidly, which leads to a small contraction of the unit-cell volume (up to $\sim$0.2 \%).

\begin{table}[b]
\caption{Atomic coordinates for Re$_{3}$Ge$_{7-x}$Ga$_{x}$.}
\renewcommand\arraystretch{1.3}
\begin{tabular}{p{1.5cm}<{\centering}p{1.2cm}<{\centering}p{0.9cm}<{\centering}p{0.9cm}<{\centering}p{0.9cm}<{\centering}p{1.9cm}<{\centering}}
\\
\hline 
   Atoms & site  &  $x$  & $y$ & $z$ & Occupancy \\

\hline 
Re(1)							& 	 8$f$ 	 & 0	& 0.074 & 0.559 & 1	\\
Re(2)							& 	  4$c$ 	 & 0 & 0.48 & 0.25 & 1		\\
Ge(1)/Ga(1)				&      8$f$    & 			0 & 0.061 & 0.163 & (7-$x$)/$x$ \\
Ge(2)/Ga(2)	&      8$f$ &   0  & 0.324 & 0.137 & (7-$x$)/$x$  \\
Ge(3)/Ga(3)				&      8$f$    & 			0 & 0.353 & 0.532 & (7-$x$)/$x$ \\
Ge(4)/Ga(4)	&      4$c$ &   0  & 0.761 & 0.25 & (7-$x$)/$x$  \\
\hline
\hline 
\end{tabular}
\label{Table3}
\end{table}

\begin{figure*}
\includegraphics*[width=17cm]{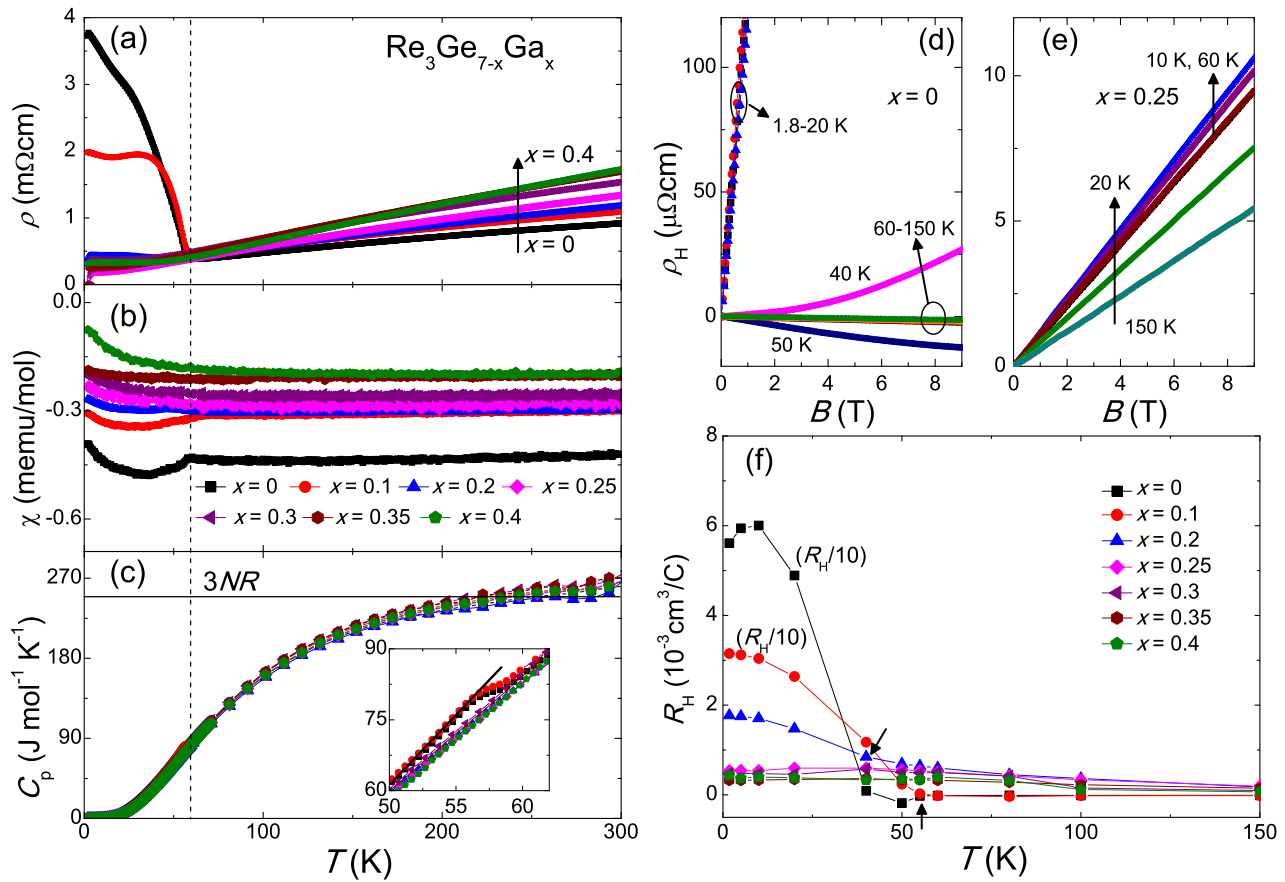}
\caption{
(a-c) Temperature dependence of resistivity, magnetic susceptibility, and specific heat below 300 K for the series of Re$_{3}$Ge$_{7-x}$Ga$_{x}$ samples, respectively. The vertical dashed line is a guide to the eyes.
In panel (a), the arrow marks the increasing of Ga content $x$. In panel (c), the horizontal line is the Dulong-Petit limit of 3$N$$R$ = 249.42 J mol$^{-1}$ K$^{-1}$, and the inset shows a zoom of the data between 50 and 62 K.
(d, e) Magnetic field dependence of Hall resistivity at various temperatures for the samples with $x$ = 0 and 0.2, respectively.
The circles and arrows are a guide to the eyes.
(f) Temperature dependence of Hall coefficient for the series of samples. The anomalies are marked by the arrows, and the data for both $x$ = 0 and 0.1 are divided by a factor of 10 for better illustration.
}
\label{fig1}
\end{figure*}
In Re$_{3}$Ge$_{7}$, Re atoms occupy two different crystallographic sites (0, 0.074, 0.559) and (0, 0.48, 0.25), and there are four distinct sites for Ge: (0, 0.061, 0.163), (0, 0.324, 0.137), (0, 0.353, 0.532), and (0, 0.761, 0.25). For the structural refinements of Re$_{3}$Ge$_{7-x}$Ga$_{x}$, the Ga atoms are assumed to be distributed randomly on the four Ge sites. Representative refinement results in the 2$\theta$ region near the strongest peak for $x$ = 0, 0.2, and 0.4 are shown in Figs. 2(a-c) [results for full patterns are shown in Fig. S1 of the Supplementary Information]. In all cases, the calculated XRD patterns match well with the observed ones, which is corroborated by the small $R_{\rm wp}$ (4.4-5.9\%) and $R_{\rm p}$ (3.2-4.3\%) factors.
Hence all samples are free of discernible impurities, confirming their high quality. In passing, we have also performed low-temperature XRD measurements on Re$_{3}$Ge$_{7}$ down to 15 K.
The data reveal that the lattice parameters vary smoothly with decreasing temperature and there is no evidence for a structural transition (see Fig. S2 of the Supplementary Information).

\begin{figure*}
\includegraphics*[width=17.8cm]{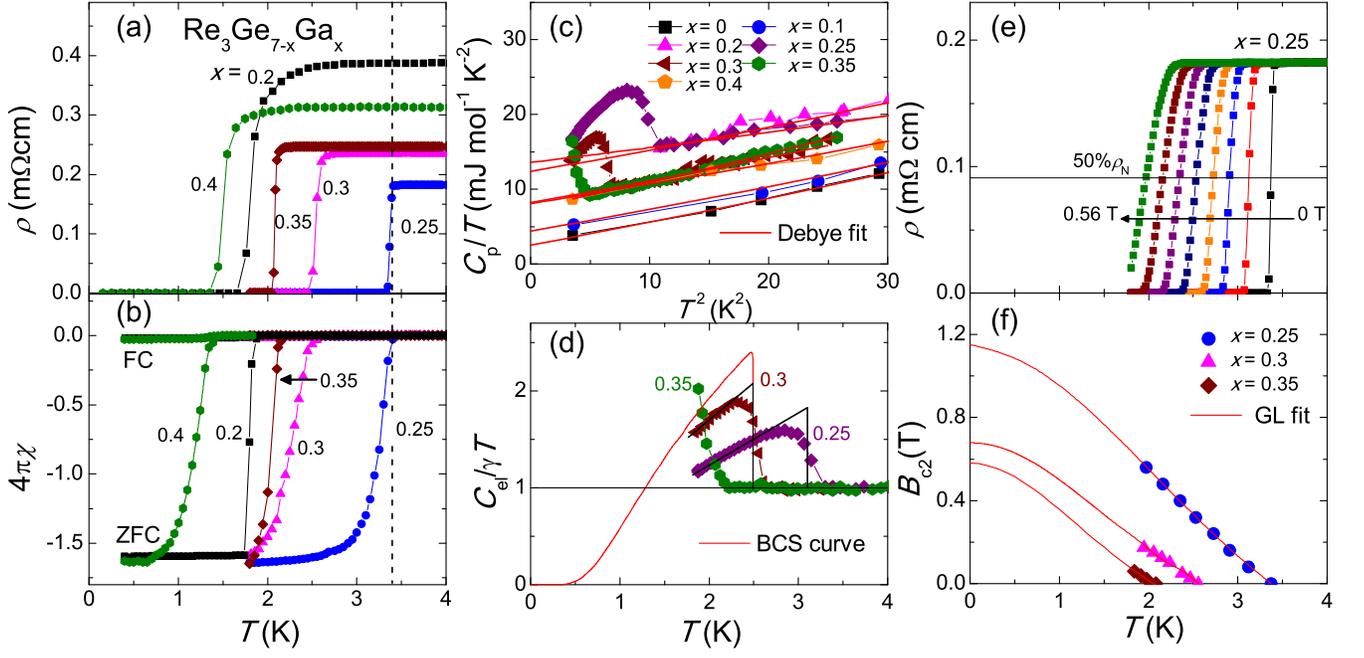}
\caption{
(a, b) Low temperature resistivity and magnetic susceptibility for Re$_{3}$Ge$_{7-x}$Ga$_{x}$ with 0.2 $\leq$ $x$ $\leq$ 0.4, respectively.
The vertical dashed line is a guide to the eyes.
(c) Low temperature specific-heat $C_{\rm p}$ data for all the Re$_{3}$Ge$_{7-x}$Ga$_{x}$ samples plotted as $C_{\rm p}$/$T$ versus $T^{2}$. The solid lines are fits by the Debye model.
(d) Temperature dependence of normalized electronic specific-heat for the samples with 0.25 $\leq$ $x$ $\leq$ 0.35. The back lines are energy conserving constructions to estimate the specific heat jump and the red line denotes the theoretical BCS curve.
(e) Temperature dependence of resistivity under various fields up to 0.56 T for the sample with $x$ = 0.2, and the field increment is 0.08 T.
The horizontal line and arrow indicate the 50\% drop of normal-state resistivity and field increasing direction, respectively.
(f) Upper critical field versus temperature phase diagram for the samples with 0.25 $\leq$ $x$ $\leq$ 0.35. The solid lines are fits to the data by the Ginzburg-Landau model.
}
\label{fig1}
\end{figure*}

A systematic change in the resistivity ($\rho$) of Re$_{3}$Ge$_{7-x}$Ga$_{x}$ is observed with increasing $x$, which is shown in Fig. 3(a).
On cooling below room temperature, the pristine Re$_{3}$Ge$_{7}$ exhibits a metallic behavior but undergoes a metal-insulator transition below $T_{\rm MI}$ = 57.3 K. At 1.8 K, the residual resistivity ratio (RRR) is much smaller than unity. This overall behavior is well consistent with that observed in single-crystal samples \cite{Re3Ge7PRM}.
When Ga is introduced into the system, the $\rho$ at high temperature increases monotonically and the metal-insulator transition is gradually suppressed.
Actually, the $\rho$ value at 1.8 K for $x$ = 0.1 is only about half that for $x$ = 0.
In addition, the application of magnetic field has little effect on the $T_{\rm MI}$ while results in a positive magnetoresistance at low temperature.
As $x$ increases above 0.2, $T_{\rm MI}$ can no longer be resolved and RRR becomes significantly larger than one with a maximum value of 7.4 at $x$ = 0.25, signifying a good metallic behavior.

The doping evolution of temperature dependent magnetic susceptibility $\chi(T)$ measured under a field of 7 T for Re$_{3}$Ge$_{7-x}$Ga$_{x}$ is displayed in Fig. 3(b).
All the $\chi(T)$ data are negative and nearly temperature independent, substantiating that the magnetic properties of these samples are dominated by the core diamagnetism. With increasing $x$, however, the $\chi(T)$ data become less negative, and hence the Pauli paramagnetic contribution seems to grow with the Ga content.
This trend implies an enhancement of the density of states at the Fermi level [$N$($E_{\rm F}$)] induced by Ga doping, consistent with the results shown below.
For $x$ $\leq$ 0.2, a drop in $\chi(T)$ is detected and its magnitude decreases as the increase of $x$. Note that the onset temperature of this anomaly agrees with $T_{\rm MI}$ determined from resistivity measurements (see the dashed line). Across the metal-insulator-like transition, it is reasonable to speculate that a gap opens at the Fermi level and hence $N$($E_{\rm F}$) decreases \cite{CDW}.
Since the Pauli paramagentic susceptibility is proportional to $N$($E_{\rm F}$), this leads to a decrease in $\chi$ as observed experimentally.
At low temperature, the $\chi(T)$ data exhibit a weak upturn, which is ascribed to a small amount of paramagnetic impurities.

\begin{table*}
\caption{Structural refinement results and physical parameters of Re$_{3}$Ge$_{7-x}$Ga$_{x}$.}
\renewcommand\arraystretch{1.3}
\begin{tabular}{p{3.4cm}<{\centering}p{1.8cm}<{\centering}p{1.8cm}<{\centering}p{1.8cm}<{\centering}p{1.8cm}<{\centering}p{1.8cm}<{\centering}p{1.8cm}<{\centering}p{1.8cm}<{\centering}p{1cm}}
\\
\hline 
   Parameter  &  $x$ = 0  & $x$ = 0.1  & $x$ = 0.2 & $x$ = 0.25 & $x$ = 0.3 & $x$ = 0.35 & $x$ = 0.4\\
\hline 
$a$ ({\AA})& 3.226(1) & 	3.227(1)  	 & 3.228(1) & 3.229(1) & 3.230(1) & 3.231(1) & 3.232(1)	\\
$b$ ({\AA})& 9.044(1) & 	9.046(1)  	 & 9.052(1) & 9.056(1) & 9.059(1) & 9.062(1) & 9.068(1)	\\
$c$ ({\AA}) & 21.960(1) 	&  21.951(1)      & 21.908(1) & 21.895(1) & 21.870(1) & 21.843(1) & 21.823(1)  \\
$R_{\rm wp}$ & 5.0\% &	5.5\%  	 & 7.5\% & 6.0\% & 5.9\% & 4.5\% & 4.6\%	\\
$R_{\rm p}$ & 3.7\% &	4.0\% 	 & 5.3\% & 4.4\%& 4.4\% & 3.3\% & 3.4\%	\\
$T_{\rm MI}$ (K) & 57.3  & 53.4   	 & 47.7 & $-$ & $-$ & $-$ & $-$	\\
$T_{\rm c}$ (K) &$-$  & $-$  	 & 1.81 & 3.37 & 2.56 & 2.08 & 1.49	\\
$\gamma$ (mJ mol$^{-1}$ K$^{-2}$) & 2.5& 	4.4  & 12.4 & 13.6 & 8.2 & 7.5 & 8.1	\\
$\Theta_{\rm D}$ (K) & 402 & 	411  	 & 416 & 468 & 411 & 389 & 427	\\
$B_{\rm c2}$(0) (T)& $-$ & $-$  	 & $-$ & 1.15 & 0.68 & 0.58& $-$	\\
$\xi_{\rm GL}$ (nm)& $-$ & 	$-$  	 & $-$ & 16.9 & 22.0 & 23.8 & $-$\\
\hline
\hline 
\end{tabular}
\label{Table3}
\end{table*}
The specific heats $C_{\rm p}$ of Re$_{3}$Ge$_{7-x}$Ga$_{x}$ are also measured and shown in Fig. 3(c).
The data for all $x$ values almost overlap with each other. In addition, the $C_{\rm p}$ values at high temperatures are close to the Dulong-Petit limit of 3$N$$R$ = 249.92 J mol$^{-1}$ K$^{-1}$, where $N$ = 10 and $R$ = 8.314 J mol$^{-1}$ K$^{-1}$ is the molar gas constant.
On close examination, a $C_{\rm p}$ anomaly is found near 57 K for $x$ = 0 and 0.1 (see the inset), confirming that the metal-insulator transition is of second order. At higher $x$ values, no such anomaly is discernible, suggesting that the transition is either too weak to be detected or completely suppressed.

The effect of Ga doping on the carrier concentration of Re$_{3}$Ge$_{7-x}$Ga$_{x}$ is further investigated by the Hall effect measurements.
Figs. 3(d) and (e) show the magnetic field dependence of Hall resistivity $\rho_{\rm H}$ for two cases of $x$ = 0 and 0.25, which show clear contrast.
In the former case, the $\rho_{\rm H}$ data are negative above 50 K, but become positive at lower temperatures.
This indicate the presence of both hole and electron carriers in the system,
which naturally explains the nonlinear $\rho_{\rm H}$ as a function of magnetic field.
In the latter case, however, $\rho_{\rm H}$ is positive and depends linearly on the field.
The temperature dependence of Hall coefficient $R_{\rm H}$ for the series of samples shown in Fig. 3(f).
Here $R_{\rm H}$ is determined as $R_{\rm H}$ = $\rho_{\rm H}$/$B$ in the low-field region.
With decreasing temperature, $R_{\rm H}$ for $x$ = 0 and 0.1 undergoes a sign reversal from negative to positive, and then rises steeply before reaching a plateau.
Note that the sign-reversal temperature is close to $T_{\rm MI}$, suggesting that the two phenomena are intimately related. At higher $x$ values, $R_{\rm H}$ is always positive and no sign-reversal occurs. Nevertheless, a rise in $R_{\rm H}$ is still observable below a temperature around $T_{\rm MI}$ for $x$ = 0.2.
Remarkably, the low-temperature $R_{\rm H}$ values for $x$ $\leq$ 0.1 are about two-orders magnitude higher than those for $x$ $\geq$ 0.25.
This implies that the latter has a much higher hole concentration, which is estimated to be $\sim$1.2-2 $\times$ 10$^{21}$ cm$^{-3}$ assuming a one-band model.
These results demonstrate that Ga doping introduces holes in Re$_{3}$Ge$_{7}$.

As a consequence of this doping, superconductivity is induced in Re$_{3}$Ge$_{7-x}$Ga$_{x}$ over an $x$ range of 0.2 to 0.4.
This is demonstrated by the $\rho$($T$) and $\chi(T)$ data below 4 K in Figs. 4(a) and (b).
As can be seen in Fig. 4(a), a drop to zero resistivity is observed for all $x$ values in this range.
With increasing $x$, the resistive transition first shifts to higher temperatures and then to lower temperatures, displaying a nonmonotonic behavior.
The $T_{\rm c}$ values, determined from the midpoints of the $\rho$ drops, are 1.81 K, 3.37 K, 2.56 K, 2.08 K, and 1.49 K for $x$ = 0.2, 0.25, 0.3, 0.35, and 0.4, respectively.
Meanwhile, $\chi_{\rm ZFC}$ of these samples measured under 1 mT exhibits a strong diamagnetic response, whose onset temperature coincides with $T_{\rm c}$.
In addition, their shielding factions are estimated to exceed $\sim$150\% without demagnetization correction.

The bulk nature of superconductivity is confirmed by the plots of low-temperature $C_{\rm p}$/$T$ versus $T^{2}$ in Fig. 4(c).
A clear $C_{\rm p}$ jump is observed for the $x$ value of 0.25, 0.3
and 0.35. As for $x$ = 0.2 and 0.4, the absence of such an anomaly is due to that their $T_{\rm c}$ values are below the lowest measurement temperature (1.8 K). On the other hand, the normal-state $C_{\rm p}$ data for both undoped and Ga-doped Re$_{3}$Ge$_{7}$ are well fitted by the Debye model
\begin{equation}
C_{\rm p}/T = \gamma + \beta T^{2} + \delta T^{4},
\end{equation}
where $\gamma$ and $\beta$($\delta$) are the electronic and phonon specific-heat coefficients, respectively. From $\beta$, the Debye temperature $\Theta_{\rm D}$ is calculated as
\begin{equation}
\Theta_{\rm D} = (\frac{12\pi^{4}NR}{5\beta})^{1/3}.
\end{equation}
The obtained $\gamma$ and $\Theta_{\rm D}$ are listed in Table II.
Intriguingly, while no systematics in $\Theta_{\rm D}$ are observed, $\gamma$ exhibits a nonmonotontic $x$ dependence with a maximum value of 13.6 mJ mol$^{-1}$ K$^{-2}$ at $x$ = 0.25.
This value is larger than those of the conventional metals \cite{conventionalmetal} and comparable to those of some iron-based superconductors with a similar carrier concentration \cite{ironSCreview}.

It is prudent to note that the $\gamma$ vaule (= 2.5 mJ mol$^{-1}$ K$^{-2}$) of our polycrystalline Re$_{3}$Ge$_{7}$ sample is much larger than that ($\sim$4 $\mu$J mol$^{-1}$ K$^{-2}$) of the single crystalline one \cite{Re3Ge7PRM}.
Moreover, the signs of their $R_{\rm H}$ data at low temperature are opposite.
These contrasts are most probably due to slight difference in stoichiometry of the two samples.
As a matter of fact, we have also grown Re$_{3}$Ge$_{7}$ crystals with (00$l$) orientation using a flux method different from that in Ref. \cite{Re3Ge7PRM}, where crystals with (0$l$0) orientation were obtained. The Hall measurements indicate that its $R_{\rm H}$ is positive in the whole temperature range (data not shown). It is thus possible that the sign of $R_{\rm H}$ depends on the crystal orientations. Since our polycrystalline samples consist of many small crystals with random orientations, it is no wonder that the sign of $R_{\rm H}$ is different from that measured on single crystals.

\begin{figure*}
\includegraphics*[width=17cm]{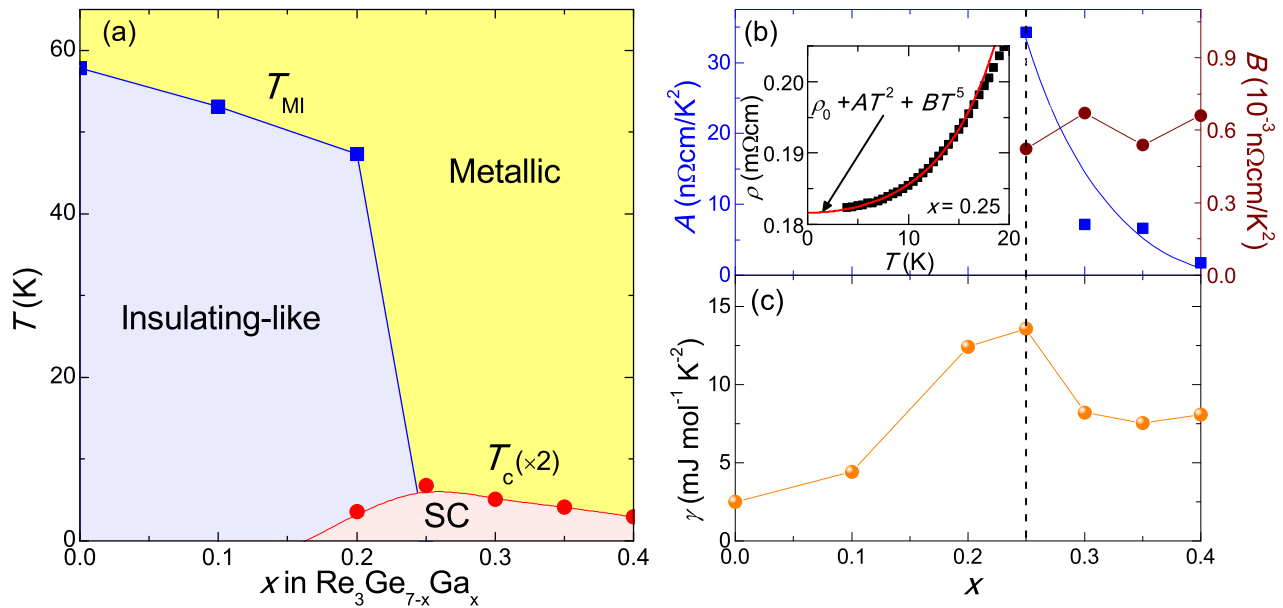}
\caption{
(a) $T$-$x$ electronic phase diagram of Re$_{3}$Ge$_{7-x}$Ga$_{x}$. Note that the $T_{\rm c}$ data are multiplied by a factor of 2 for clarity. (b, c) $x$ dependence of the resistivity fitting parameters (0.25 $\leq$ $x$ $\leq$ 0.4) and electronic specific-heat coefficient, respectively. The inset of (b) shows, as an example, the fitting of the normal-state resistivity data for $x$ = 0.25 by the power law $\rho$ = $\rho_{0}$ + $A$$T^{2}$ + $B$$T^{5}$ (see text for details). The vertical dashed line is a guide to the eyes.
}
\label{fig1}
\end{figure*}
The normalized electronic specific heat $C_{\rm el}$/$\gamma$$T$ for 0.25 $\leq$ $x$ $\leq$ 0.35, obtained by subtraction of the phonon contribution, is shown in Fig. 4(d).
It turns out that the $C_{\rm el}$/$\gamma$$T$ jump decreases with increasing $x$ and hence increasing $T_{\rm c}$.
Using an entropy conserving construction, $\Delta$$C_{\rm el}$/$\gamma$$T$ is determined to be 0.83 and 1.08 for $x$ = 0.25 and 0.3, respectively, which are significantly smaller than the BCS value of 1.43 \cite{BCSthoery}. Indeed, the temperature dependence of their $C_{\rm el}$/$\gamma$$T$ data show a clear deviation from the weak coupling BCS theory \cite{BCSthoery}, hinting at the presence of multiple superconducting gaps or even gap nodes. Hence, to better understand the gap structure, $C_{\rm p}$ measurements at temperatures much below $T_{\rm c}$ are needed. Nevertheless, since inhomogeneity could be present is our polycrystalline samples, such investigation is best performed on single crystals and thus left for future studies.

The upper critical fields $B_{\rm c2}$ for Re$_{3}$Ge$_{7-x}$Ga$_{x}$ with 0.25 $\leq$ $x$ $\leq$ 0.35 are determined by resistivity measurements under magnetic fields.
An example for $x$ = 0.25 is shown in Fig. 4(e). As expected, the resistive superconducting transition shifts toward lower temperatures and becomes broadened as the field increases.
At each field, the $T_{\rm c}$ value is determined using the same criterion as above.
The resulting $B_{\rm c2}$ versus temperature phase diagrams are displayed in Fig. 4(f).
All the $B_{\rm c2}$($T$) data are well described by the Ginzburg-Landau (GL) model
\begin{equation}
B_{\rm c2}(T) = B_{\rm c2}(0)\frac{1-t^{2}}{1+t^{2}},
\end{equation}
where $B_{\rm c2}(0)$ is the zero-temperature upper critical field and $t$ = $T$/$T_{\rm c}$.
The obtained $B_{\rm c2}$(0) is 1.15 T for $x$ = 0.25, 0.68 T for $x$ = 0.3, and 0.58 T for $x$ = 0.35.
Once $B_{\rm c2}$(0) is known, the GL coherence length $\xi_{\rm GL}$ can be calculated by the equation
\begin{equation}
\xi_{\rm GL} = \sqrt{\frac{\Phi_{0}}{2\pi B_{\rm c2}(0)}},
\end{equation}
where $\Phi_{0}$ = 2.07 $\times$ 10$^{-15}$ Wb is the flux quantum.
This gives $\xi_{\rm GL}$ values of 16.9 nm, 22.0 nm and 23.8 nm for $x$ = 0.25, 0.3 and 0.35, respectively.
\begin{figure*}
\includegraphics*[width=17.5cm]{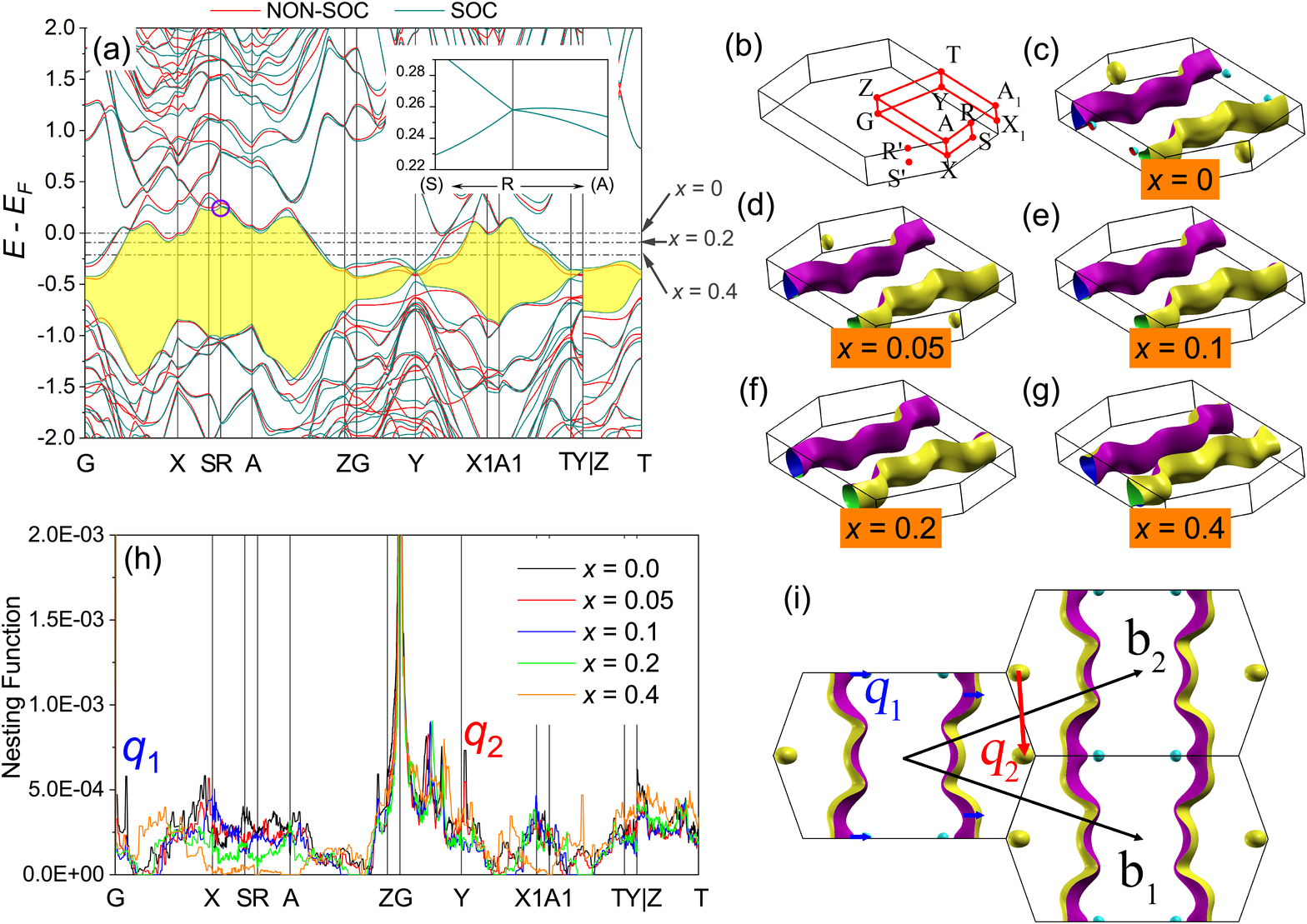}
\caption{
(a) Bulk band structure for Re$_{3}$Ge$_{7-x}$Ga$_{x}$ without and with SOC. The yellow shaded region corresponds to the band filling of $x~=~1$. (b) Brillouin zone of Re$_{3}$Ge$_{7-x}$Ga$_{x}$ with high symmetry points and lines. (c-g) The Fermi surface for Re$_{3}$Ge$_{7-x}$Ga$_{x}$ with $x$ = 0, 0.05, 0.1, 0.2 and 0.4, respectively. (h) The calculated Fermi surface nesting functions at high-symmetry lines. (i) A diagrammatic sketch of nesting vectors $q_{1}$ and $q_{2}$.}
\label{fig6}
\end{figure*}

The above results, which are summarized in Table II, allow us to construct the electronic phase diagram of Re$_{3}$Ge$_{7-x}$Ga$_{x}$ presented in Fig. 5(a).
Upon Ga doping, $T_{\rm MI}$ in Re$_{3}$Ge$_{7}$ is gradually suppressed and disappears abruptly at $x$ $>$ 0.2. Note that, below $T_{\rm MI}$, the carrier concentration decreases by more than one order of magnitude as indicated by the above Hall measurements. This is not typical for a semimetal but reminiscent of an insulating-like behavior, as we labeled in the phase diagram. On the other hand, superconductivity emerges for $x$ $\geq$ 0.2, and $T_{\rm c}$ shows a dome-like dependence on $x$ with a maximum of 3.37 K observed at $x$ = 0.25. The contrasting behavior of $T_{\rm MI}$ and $T_{\rm c}$ suggests that there is a competition between the insulating-like and superconducting phases, though the two phases might coexist in a narrow $x$ range between 0.2 and 0.25. It should be pointed out, for each $x$ value, we have carried out measurements on multiple samples.
An example of $x$ = 0.25 and a summarizing table are shown in S3 and S4, respectively, of the Supplementary Information. These results indicate that the error bar in characteristic temperatures is within the symbol size, and hence the overall phase diagram is well reproducible.

In order to find clues to superconducting mechanism, the normal-state $\rho$($T$) data are fitted by the formula
\begin{equation}
\rho(T) = \rho_{0} + AT^{2} +BT^{5},
\end{equation}
where $\rho_{0}$ is the residual resistivity, and $T^{2}$ and $T^{5}$ terms are the contributions from electron-electron and electron-phonon scattering, respectively. Here we restrict ourself to $x$ $\geq$ 0.25 to avoid influence from the metal-insulator-like transition. The $x$ dependencies of prefactors $A$ and $B$ are displayed in Fig. 5(b) and an example of the fitting for $x$ = 0.25 is shown in the inset.
With increasing $x$ in this range, while the prefactor $B$ is essentially $x$ independent, the prefactor $A$ drops by nearly one order of magnitude.
Although the situation is unclear at lower $x$ values, it is reasonable to speculate that $A$ reaches a maximum around $x$ = 0.25.
This is also supported by the observation of a maximal $\gamma$ at this $x$ value, as can be seen from Fig. 4(c).
Remarkably, with the maximal values of $A$ and $\gamma$, one finds a Kadowaki-Woods ratio $A$/$\gamma^{2}$ of 1.7 $\times$ 10$^{-4}$ $\mu\Omega$ cm mol$^{2}$ K$^{2}$ mJ$^{-2}$. This value is considerably larger than that of heavy fermions (1.0 $\times$ 10$^{-5}$ $\mu\Omega$ cm mol$^{2}$ K$^{2}$ mJ$^{-2}$) \cite{KW}, suggesting significant electron correlation in the present system.
Furthermore, as the system moves away from optimal doping, electron-electron scattering is strongly weakened but the electron-phonon scattering strength remains little affected. These results point to the important role of electron-electron interaction played in Cooper pairing.
\begin{figure*}
\includegraphics*[width=17.5cm]{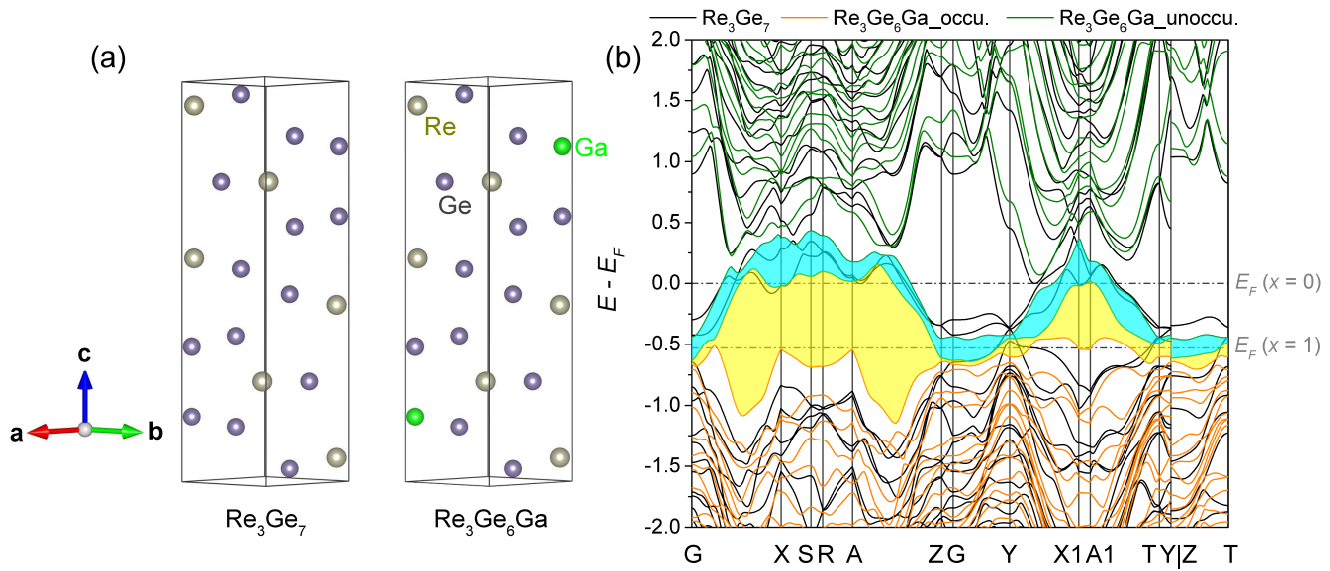}
\caption{
(a) The primitive cell for Re$_{3}$Ge$_7$ and a possible $C2/m$ configuration of Re$_{3}$Ge$_6$Ga. (b) Bulk band structure for Re$_3$Ge$_6$Ga with space group $C2/m$. The cyan and yellow regions correspond to the band filling of $x$ = 0 and 1, respectively.}
\label{fig7}
\end{figure*}

To gain insight into the effect of Ga doping in Re$_{3}$Ge$_{7}$, we performed first-principles calculations on the electronic structure of Re$_{3}$Ge$_{7-x}$Ga$_{x}$ with $x$ up to 0.4.
The calculated band dispersions with and without spin-orbit coupling (SOC) are shown in Fig. 6(a), and the Brillouin zone is sketched in Fig. 6(b).
One can see that there are several bands crossing $E_{\rm F}$ and the Ga doping mainly leads to a rigid band shift of the Fermi level ($E_{\rm F}$) (see Fig. S5 of the Supplementary Information).
Without considering SOC, there are three 4-fold degenerate points about 0.3 eV above $E_{\rm F}$ along the X-R line.
Nevertheless, when turning on SOC, only the one located at R point remains ungapped. Group theory analysis of the \emph{\textbf{k}$~\cdot$~\textbf{p}} perturbation matrix elements shows that the energy bands around this 4-fold point have linear dispersions \cite{Dirac-PRL, Dirac-thesis}, which is protected by the nonsymmorphic space group $Cmcm$. Since R is a time-reversal invariant momentum, the linear dispersions for Kramers partners must be reversed. Thus despite that the band dispersion is tilted, the 4-fold point at R is a type-I Dirac point \cite{Dirac-review}, in agreement with the previous report \cite{Re3Ge7topology1}.

Figure 6(c-h) shows the calculated Fermi-surface and its nesting function along the high symmetry lines for Re$_{3}$Ge$_{7-x}$Ga$_{x}$.
For undoped Re$_{3}$Ge$_{7}$ ($x$ = 0), its Fermi surface consists of two large hole pockets along the $\textbf{k}_y$ direction and two sets of small electron pockets crossing the G-X and Y-X1 lines.
Upon Ga doping, the hole pockets remain nearly unchanged while the electron pockets become smaller and vanish for $x$ $\geq$ 0.1.
These results are well consistent with the evolution of $R_{\rm H}$ data and confirm Ga as a hole dopant.
Nevertheless, since the metal-insulator-like transition is not considered in the calculation, a direct comparison between the evolutions of theoretical density-of-states (DOS) and experimental $\gamma$ values with Ga content is not meaningful.
On the other hand, the nesting function exhibits peaks around X, X1, Z points and along G-X, G-Y-X1 lines, in addition to the universal one at $\Gamma$ point.
Nevertheless, only the peaks along G-X and Y-X1 lines (marked by $q_1$ and $q_2$ in Fig. 6(h)) are significantly suppressed with Ga doping, while the other peaks survive at higher doping level.
The corresponding wave vectors for these peaks are $q_1~=~$($\pm$0.06, 0, 0) and $q_2~=~$($\pm$0.02, 1, 0) in the unit of ($\pi/a$, $\pi/b$, $\pi/c$). By careful inspection of the Fermi surface structure, we find that the nesting peaks at $q_1$ and $q_2$ are closely related to the electron pockets, as shown in Fig. 6(i). The $q_1$ peak should be mainly contributed by the nesting between the hole and electron pockets. Nevertheless, since trace of $q_1$ is still present at higher Ga doping levels, it is reasonable to infer that this peak is also contributed by the nesting between the hole pockets.
For the peak of $q_2$, it disappears as soon as the electron pockets vanish and thus should result from the nesting between the ellipsoid-shaped electron pockets around X point. This is corroborated by the 2D nesting functions shown in Fig. S6 of the Supplementary Information.
It is noted that the metal-insulator-like transition and the above-mentioned nesting peaks are suppressed at almost the same Ga doping level. It is therefore reasonable to speculate that the phase transition in pristine Re$_{3}$Ge$_{7}$ is associated with a nesting driven Fermi surface instability \cite{FSnesting-1,FSnesting-2}.

Now let's examine the band topology of Re$_{3}$Ge$_{7-x}$Ga$_{x}$.
As illustrated in Fig. 6(a), a moderate doping of $x~=~$1 would lower the electron filling level into the yellow full-gapped region.
Given that Re$_3$Ge$_7$ holds the spatial inversion symmetry, one can obtain the $Z_2$ invariants by calculating the inversion parities \cite{3DTI, 3DTI_inv}.
This gives $Z_2$ invariants of (1;110), which corresponds to a strong topological insulator (TI) phase.
Since the Dirac point is protected by the nonsymmorphic space group, it is of interest to investigate the configuration that breaks the nonsymmorphic group symmetry.
In Fig. 7(a), we present a hypothetical structure of Re$_3$Ge$_6$Ga, which is derived by replacing 2 Ge atoms of Ge(1) site with Ga (per primitive cell). Such a configuration lowers the symmetry group to $C2/m$, which has only 2-dimensional double valued irreducible representations. Thus one expect that the bands are doubly degenerated at any high-symmetry points, which split the Dirac point into two doubly degenerated points. The calculated band structure for Re$_3$Ge$_6$Ga is shown in Fig. 7(b). Indeed, the Dirac point at R point splits into a full-gapped region (the cyan shaded region), while the yellow shaded region remains gapped. The $Z_2$ invariants are calculated to be (0;001) and (1;110) for the cyan and yellow regions, respectively. This not only demonstrates the robustness of the strong TI phase in Ga-doped Re$_{3}$Ge$_{7}$, but also indicates that Re$_3$Ge$_7$ will become a weak TI by breaking the nonsymmorphic group symmetry.

Finally, we discuss the implications of our results and suggest some directions for future research.
First, since the overall properties of Re$_{3}$Ge$_{7-x}$Ga$_{x}$ are similar to those of Cu$_{x}$TiSe$_{2}$ \cite{CDW}, the phase transition in Re$_{3}$Ge$_{7}$ could be due to the CDW formation, which calls for verification by temperature-dependent electron diffraction studies.
Second, the emergence of superconductivity by only $\sim$3\% Ga doping implies that Re$_{3}$Ge$_{7}$ lies on the verge of a superconducting instability. It is therefore of interest to see whether superconductivity can be induced by doping with other elements or the application of high pressure.
Third, the combination of superconductivity and nontrivial band topology renders Re$_{3}$Ge$_{7-x}$Ga$_{x}$ a potential candidate for TSC \cite{TSCreview1,TSCreview2,TSCreview3}.
In this respect, single crystal growth is highly desirable for further spectroscopy measurements on the superconducting gap symmetry and possible in-gap states.

\section{III. Conclusions}
In summary, we have discovered superconductivity in the orthorhombic Re$_{3}$Ge$_{7-x}$Ga$_{x}$ system.
The pristine Re$_{3}$Ge$_{7}$ exhibits a second order metal-insulator-like phase transition below 57.3 K, which is suppressed upon Ga doping and disappears above $x$ = 0.2. At this doping level, superconductivity starts to be observed and $T_{c}$ displays a dome-like dependence with a maximum value of 3.37 K at $x$ = 0.25.
The Hall effect measurements indicate that substitution of Ga for Ge introduces holes, consistent with the band structure calculations.
The theoretical results further suggest that the phase transition in Re$_{3}$Ge$_{7}$ is likely driven by Fermi surface nesting and superconducting Re$_{3}$Ge$_{7-x}$Ga$_{x}$ compositions exhibit nontrivial band topology characterized by strong $Z_{2}$ invariants.
Our results indicate that Ga doped Re$_{3}$Ge$_{7}$ is a rare system that combines collective phase transition, nontrivial band topology and superconductivity, which lays a foundation for further exploring the competition and interplay between these properties.

\section{IV. Method}
\textbf{Sample synthesis.} Polycrystalline Re$_{3}$Ge$_{7-x}$Ga$_{x}$ samples with $x$ = 0, 0.1, 0.2, 0.25, 0.3, 0.35 and 0.4 were prepared by the solid-state reaction method. High-purity Re (99.99\%), Ge (99.99\%) powders, and Ga (99.999\%) shots were weighed according to the stoichiometric ratio, mixed thoroughly and pressed into pellets in an argon-filled glove box. The pellets were then sealed in evacuated silica tubes and heated at 850 $^{\circ}$C for several days, followed by slow cooling to room temperature. This process was repeated several times with intermediate grindings to ensure homogeneity.

\textbf{Structural and chemical characterizations.} The phase purity of resulting samples was examined by powder x-ray diffraction (XRD) using a Bruker D8 Advance x-ray diffractometer with Cu K$\alpha$ radiation.
The data were collected with a step-scan mode in the 2$\theta$ range from 5$^{\circ}$ to 120$^{\circ}$
and the structural refinements were performed using the program JANA2006 \cite{JANA}. The sample morphology and chemical composition were characterized by a Zeiss Supratm 55 schottky field emission scanning electron microscope (SEM) with an energy-dispersive x-ray (EDX) spectrometer.

\textbf{Physical property measurements.} Measurements of resistivity, Hall coefficient and specific heat were done on regular-shaped samples in a Quantum Design Physical Property Measurement System (PPMS-9 Dynacool).
The resistivity was measured by the four-probe method and down to 150 mK at zero field using an adiabatic dilution refrigerator option. The Hall resistivity was measured by sweeping the field from $-$9 T to 9 T, and the data were antisymmetrized to remove the magnetoresistance contribution. The zero-field cooling (ZFC) and field cooling (FC) magnetic susceptibility measurements down to 0.4 K were carried out using a Quantum Design Magnetic Property Measurement System (MPMS3).

\textbf{Theoretical calculations.} Our first-principles calculations were performed within density functional theory (DFT), as implemented in the Vienna Ab-initio Simulation Package (VASP) \cite{VASP}. The Kohn-Sham equations were constructed on a projector augmented wave (PAW) basis \cite{PAW}. The exchange-correlation energy was calculated with a Perdew-Burke-Ernzerhof (PBE) type functional \cite{GGA}. For all calculations, we adopted a monoclinic primitive cell with experimental lattice parameters. The plane-wave energy cutoff for wavefunctions was set to 600 eV. The \emph{\textbf{k}}-mesh was set 12$\times$12$\times$3 for self-consistent calculations and 24$\times$24$\times$6 for DOS calculations.
To calculate the band dispersion over the whole Brillouin zone more efficiently, we constructed a tight-binding Hamiltonian with maximally localized Wannier functions (MLWF) \cite{MLMF}. The Fermi surfaces and nesting functions were then calculated with this tight-binding Hamiltonian.

\section*{ACKNOWLEGEMENT}
We acknowledge financial support by the foundation of Westlake University and National Key Research Development Program of China (No.2017YFA0303002).

\section*{Competing interests}
The authors declare no competing interests.

\section*{Author Contributions}
Y.W.C. and Z.R. conceived the project. Y.W.C. synthesized the samples and did the physical property measurements with the assistance from Q.Q.Z, G.X.X., B.L. and J.F.W.. S.Q.W. and G.H.C. performed theoretical calculations. RZ supervised the project and wrote the paper with inputs from Y.W.C. and S.Q.W..

\section*{Data availability}
The data that support the findings of this study are available from the corresponding author upon reasonable request.


\begin{thebibliography}{00}

\bibitem{cuprate}
J. G. Bednorz and K. A. Muller.
Possilbe high $T_{\rm c}$ superconductivity in the Ba-La-Cu-O system.
Z. Phys. B {\bf 64}, 189 (1986).

\bibitem{HF}
D. Jaccard, K. Behnia, and J. Sierro.
Pressure induced heavy fermion superconductivity of CeCu$_{2}$Ge$_{2}$.
Phys. Lett. A {\bf 163}, 475 (1992).

\bibitem{CDW}
E. Morosan, H. W. Zandbergen, B. S. Dennis, J. W. G. Bos, Y. Onose, T. Klimczuk, A. P. Ramirez, N. P. Ong, and R. J. Cava.
Superconductivity in Cu$_{x}$TiSe$_{2}$.
Nat. Phys. {\bf 2}, 544 (2006).

\bibitem{ironpnictide}
Y. Kamihara, T. Watanabe, M. Hirano, and H. Hosono.
Iron-Based Layered Superconductor La[O$_{1-x}$F$_{x}$]FeAs ($x$ = 0.05-0.12) with $T_{\rm c}$ = 26 K.
J. Am. Chem. Soc. {\bf 130}, 3296 (2008).

\bibitem{parent1}
J. I. Budnick, A. Golnik, Ch. Nieddermayer, E. Recknagel, M. Rossmanith, A. Weidinger, B. Chamberland, M. Filipkowski, and D. P. Yang.
Observation of magnetic ordering in La$_{2}$CuO$_{4}$ by moun spin rotation spectroscopy.
Phys. Lett. A {\bf 124}, 103 (1987).

\bibitem{parent2}
F. R. de Boer, J. C. P. Klaasse, P. A. Veenhuizen, A. Bohm, C. D. Bredl, U. Gottwick, H. M. Mayer, L. Pawlak, U. Rauchschwalbe, H. Spille, and F. Steglich.
CeCu$_{2}$Ge$_{2}$: Magnetic order in a Kondo lattice.
J. Magn. Magn. Mater. {\bf 63}, 91 (1987).

\bibitem{parent3}
C. de la Cruz, Q. Huang, J. W. Lynn, J. Y. Li, W. Ratcliff II, J. L. Zarestky, H. A. Mook, G. F. Chen, J. L. Luo, N. L. Wang, and P. C. Dai.
Magnetic order close to superconductivity in the iron-based layered LaO$_{1-x}$F$_{x}$FeAs systems.
Nature {\bf 453}, 899 (2008).

\bibitem{parent4}
F. J. Di Salvo, D. E. Moncton, and J. V. Waszczak.
Electronic properties and superlattice formation in the semimetal TiSe$_{2}$.
Phys. Rev. B {\bf 14}, 4321 (1976).

\bibitem{MT1}
M. A. McGuire, A. D. Christianson, A. S. Sefat, B. C. Sales, M. D. Lumsden, R. Y. Jin, R. A. Payzant, D. Mandrus, Y. B. Luan, V. Keppens, V. Varadarajan, J. W. Brill, R. P. Hermann, M. T. Sougrati, F. Grandjean, and G. J. Long.
Phase transitions in LaFeAsO: Structural, magnetic, elastic, and transport properties, heat capacity and Mossbauer spectra.
Phys. Rev. B {\bf 78}, 094517 (2008).

\bibitem{MT2}
D. J. Campbell, C. Eckberg, P. Y. Zavalij, H. H. Kung, E. Rzaaoli, N. Michiardi C. Jozwiak, A. Bostwick, E. Rotenberg, A. Damascelli, and J. Paglione.
Intrinsic insulating ground state in transition metal dichalcogenide TiSe$_{2}$.
Phys. Rev. Materials {\bf 3}, 053402 (2019).

\bibitem{TSCreview1}
X. L. Qi and S. C. Zhang.
Topological insulators and superconductors.
Rev. Mod. Phys. {\bf 83}, 1057 (2011).

\bibitem{TSCreview2}
J. Alicea.
New directions in the pursuit of Majorana fermions in solid state systems.
Rep. Prog. Phys. {\bf 75}, 076501 (2012).

\bibitem{TSCreview3}
M. Sato and Y. Ando.
Topological superconductors: a review.
Rep. Prog. Phys. {\bf 80}, 076501 (2017).

\bibitem{doping1}
Y. S. Hor, A. J. Williams, J. G. Checkelsky, P. Roushan, J. Seo, Q. Xu, H. W. Zandbergen, A. Yazdani, N. P. Ong, and R. J. Cava.
Superconductivity in Cu$_{x}$Bi$_{2}$Se$_{3}$ and its implications for pairing in the undoped topological insulator.
Phys. Rev. Lett. {\bf 104}, 057001 (2010).

\bibitem{doping2}
S. Sasaki, Z. Ren, A. A. Taskin, K. Segawa, L. Fu, and Y. Ando.
Odd-parity pairing and topological superconductivity in a strongly spin-orbit coupled semiconductor.
Phys. Rev. Lett. {\bf 109}, 217004 (2012).

\bibitem{doping3}
Z. H. Liu, X. Yao, J. F. Shao, M. Zuo, L. Pi, S. Tan, C. J. Zhang, and Y. H. Zhang.
Superconductivity with Topological Surface State in Sr$_{x}$Bi$_{2}$Se$_{3}$.
J. Am. Chem. Soc. {\bf 137}, 10512 (2015).

\bibitem{doping4}
Z. W. Wang, A. A. Taskin, T. Frolich, M. Braden, and Y. Ando.
Superconductivity in Tl$_{0.6}$Bi$_{2}$Te$_{3}$ Derived from a Topological Insulator.
Chem. Mater. {\bf 28}, 779 (2016).

\bibitem{doping5}
T. Asaba, B. J. Lawson, C. Tinsman, L. Chen, P. Corbae, G. Li, Y. Qiu, Y. S. Hor, L. Fu, and L. Li.
Rotational Symmetry Breaking in a Trigonal Superconductor Nb-doped Bi$_{2}$Se$_{3}$.
Phys. Rev. X {\bf 7}, 011009 (2017).

\bibitem{doping6}
L. Zhu, Q. Y. Li, Y. Y. Lv, S. C. Li, X. Y. Zhu, Z. Y. Jia, Y. B. Chen, J. S. Wen, and S. C. Li.
Superconductivity in Potassium-Intercalated $T_{d}$-WTe$_{2}$.
Nano Lett. {\bf 18}, 6585 (2018).

\bibitem{doping7}
J. F. Wu, C. Q. Hua, B. Liu, Y. W. Cui, Q. Q. Zhu, G. X. Xiao, S. Q. Wu. G.-H. Cao, Y. H. Lu, and Z. Ren.
Doping-Induced Superconductivity in the Topological Semimetal Mo$_{5}$Si$_{3}$.
Chem. Mater. {\bf 32}, 8930 (2020).

\bibitem{Re3Ge7}
T. Siegrist, F. Hulliger, and W. Petter.
The crystal sructure of Re$_{3}$Ge$_{7}$.
J. Less Common Met. {\bf 90}, 143 (1983).

\bibitem{Re3Ge7PRM}
A. Rabus and E. Mun.
Anomalous transport properties of Re$_{3}$Ge$_{7}$.
Phys. Rev. Mater. {\bf 3}, 013404 (2019).

\bibitem{Re3Ge7topology1}
T. T. Zhang, Y. Jiang, Z. D. Song, H. Huang, Y. Q. He, Z. Fang, H. M. Weng, and C. Fang.
Catalogue of topological electronic materials.
Nature {\bf 566}, 475 (2019).

\bibitem{radius}
Y. Tarutani and M. Kudo.
Atomic radii and lattice parameters of the A15 crystal structure.
J. Less Common Met. {\bf 55}, 221 (1977).

\bibitem{conventionalmetal}
C. Kittel.
Introduction to Solid State Physics.
John Wiley, New York, NY, 1996.

\bibitem{ironSCreview}
G. R. Stewart.
Superconductivity in iron compounds.
Rev. Mod. Phys. {\bf 83}, 1589 (2011).

\bibitem{BCSthoery}
J. Bardeen, L. N. Cooper, and J. R. Schreiffer.
Theory of Superconductivity.
Phys. Rev. {\bf 108}, 1175 (1957).

\bibitem{KW}
K. Kadowaki and S. B. Woods.
Universal relationship of the resistivity and specific heat in heavy-Fermion compounds.
Solid State Commun. {\bf 58}, 507 (1986).

\bibitem{Dirac-PRL}
S. M. Young, S. Zaheer, J. C. Y. Teo, C. L. Kane, E. J. Mele, and A. M. Rappe.
Dirac Semimetal in Three Dimensions.
Phys. Rev. Lett. {\bf 108}, 140405 (2012).

\bibitem{Dirac-thesis}
S. Zaheer.
Three dimensional Dirac semimetals.
PhD thesis, Univ. Penn. (2014).

\bibitem{Dirac-review}
H. Gao, J. W. F. Venderbos, Y. Kim, and A. M. Rappe.
Topological semimetal from first-principles.
Annu. Rev. Mater. Res. {\bf 49}, 153 (2019).

\bibitem{FSnesting-1}
A. Landa, J. Klepeis, P. Soderlind, I. Naumov, O. Velikokhatnyi, L. Vitos, and A. Ruban.
Fermi surface nesting and pre-martensitic softening in V and Nb at high pressures.
J. Phys.: Condens. Matter {\bf 18}, 5079C5085 (2006).

\bibitem{FSnesting-2}
I. Errea , M. M. Canales , A. R. Oganov, and A. Bergara.
Fermi surface nesting and phonon instabilities in simple cubic calcium.
High Pressure Res. {\bf 28}, 443C448 (2008).

\bibitem{3DTI}
L. Fu, C. L. Kane, and E. J. Mele.
Topological Insulators in Three Dimensions.
Phys. Rev. Lett. {\bf 98}, 106803 (2007).

\bibitem{3DTI_inv}
L. Fu, and C. L. Kane.
Topological insulators with inversion symmetry.
Phys. Rev. B. {\bf 76}, 045302 (2007).

\bibitem{JANA}
V. Petricek, M, Dusek, and L. Palatinus.
Crystallographic Computing System JANA2006: General features.
Z. Kristallogr. {\bf 229}, 345 (2014).

\bibitem{VASP}
G. Kresse and J. Hafner.
\emph{Ab initio} molecular dynamics for liquid metals.
Phys. Rev. B {\bf 47}, 558 (1993).

\bibitem{PAW}
P. E. Blochl.
Projector augmented-wave method.
Phys. Rev. B {\bf 50}, 17953 (1994).

\bibitem{GGA}
J. P. Perdew, K. Burke, and M. Ernzerhof.
Generalized Gradient Approximation Made Simple.
Phys. Rev. Lett. {\bf 77}, 3865 (1996).

\bibitem{MLMF}
N. Marzari and D. Vanderbilt.
Maximally localized generalized Wannier functions for composition energy bands.
Phys. Rev. B {\bf 56}, 12847 (1997).

\end{thebibliography}
\end{document}